\documentclass[twocolumn, aps, pra]{revtex4-1}

\usepackage{amssymb}
\usepackage{graphicx}
\usepackage{natbib}

\begin{document}

\title{Metamaterial Polarization Converter Analysis: Limits of Performance}

\author{Dmitry~L.~Markovich}
\affiliation{Department of Photonics and Information Technology,
St.~Petersburg National Research University of Information Technologies, Mechanics and Optics, St. Petersburg, 197101, Russian Federation}

\author{Andrei Andryieuski}
\email[]{andra@fotonik.dtu.dk}
\affiliation{DTU Fotonik - Department
of Photonics Engineering, Technical University of Denmark,
{\O}rsteds pl. 343, DK-2800 Kongens Lyngby, Denmark}
\author{Maksim Zalkovskij}
\affiliation{DTU Fotonik - Department of Photonics Engineering,
Technical University of Denmark, {\O}rsteds pl. 343, DK-2800 Kongens
Lyngby, Denmark}
\author{Radu Malureanu}
\affiliation{DTU Fotonik - Department of Photonics Engineering,
Technical University of Denmark, {\O}rsteds pl. 343, DK-2800 Kongens
Lyngby, Denmark}
\author{Andrei~V.~Lavrinenko}
\affiliation{DTU Fotonik - Department of Photonics Engineering,
Technical University of Denmark, {\O}rsteds pl. 343, DK-2800 Kongens
Lyngby, Denmark}

\date{\today}

\begin{abstract}
In this paper we analyze the theoretical limits of a metamaterial converter that allows for linear-to-elliptical polarization transformation with any desired ellipticity and ellipse orientation. We employ the transmission line approach providing a needed level of the design generalization. Our analysis reveals that the maximal conversion efficiency for transmission through a single metamaterial layer is 50\%, while the realistic reflection configuration can give the conversion efficiency up to 90\%. We show that a double layer transmission converter and a single layer with a ground plane can have 100\% polarization conversion efficiency. We tested our conclusions numerically reaching the designated limits of efficiency using a simple metamaterial design. Our general analysis provides useful guidelines for the metamaterial polarization converter design for virtually any frequency range of the electromagnetic waves.
\end{abstract}



\maketitle

\section{Introduction}
Metamaterials provide new exciting possibilities for light wave manipulations especially with operations on the wave polarization, which are on demand not only in the optical and microwave range, but also in the booming field of terahertz (THz) science and technology, due to the natural limitations of the material properties. THz waves have high potential in communication systems, food quality control, defense, biomedical imaging and chemical spectroscopy \cite{Jepsen2011,Kleine-Ostmann2011,Tonouchi2007}. For some THz applications, for example, magneto-optical spectroscopy \cite{Molter2012}, it is desirable to have a circularly or elliptically-polarized wave, while most THz sources generate linearly polarized radiation.

There are two main routes to get the polarization rotation or conversion. The "phase" route is to introduce eigenwaves phase offsets in birefringent or gyrotropic media with approximately equal transmitted amplitudes.  The "amplitude" route is to play with transmission coefficients for the eigenwaves letting the output to have a polarization state of the dominating eigenwave. The unnecessary polarization is then discriminated by the higher absorption and/or higher reflection of another eigenstate. Two illustrative examples of these routes in the THz range are an achromatic quarter-wave plate made from quartz \cite{Masson2006} and a giant Faraday effect in an electron plasma in n-InSb semiconductor \cite{Arikawa2012}. Being broadband, these solutions, however, claim extended sizes (3 cm) of devices in the former case and intense magnetic fields of several tesla in the latter.

In contrary to the aforementioned bulk devices, the metamaterials (metasurfaces, frequency selective surfaces) based solutions can be extremely compact, for example, a single thin layer can be enough to get the required polarization state without external magnetic field. Several metamaterials-based polarization conversion devices have already been proposed, which can be tentatively grouped by the operational principle and configuration: birefringent polarizers \cite{Strikwerda2009,Saha2010}, transmission \cite{Drezet2008,Chin2008,Chin2009,Peralta2009,Weis2009,Li2010,Roberts2012,Sun2011} and reflection \cite{Hao2007,Hao2009,Pors2011,Strikwerda2011,Wang2012} polarizers based on resonant particles or slits and chiral metamaterials \cite{Singh2010,Li2011,Mutlu2011,Zhao2012,Sabah2012,Gansel2012}. Some of the proposed devices have drawbacks, for example, being based on resonant inclusions, apertures or meta-atoms they usually exhibit a narrow bandwidth or convert a linear polarization into the specific circular one. Nevertheless, such big variety of converters' designs poses the question on whether the natural bounds for the conversion efficiency exist and how it is possible to approach them from the practical point of view.

In this contribution we evaluate the theoretical limits of the efficiency of metamaterial-based polarization converters. We consider the case of conversion of a linear polarization into any elliptical one with a desired ellipticity and ellipse orientation. We employ the transmission line theory, which proved to be useful in the theory of metamaterials \cite{Caloz2006}. Two principle experimental configurations are considered: reflection and transmission at normal incidence. We show that the conversion efficiency can be virtually as high as $50\%$ for one transmission layer and up to full $100\%$ for a polarization converter with two layers. In the reflection configuration even with only one layer conversion of up to $80-90\%$ is feasible, while one layer with a ground plane can give $100\%$ conversion with the relaxed requirements on the metamaterial unit cell design. Demonstration is exemplified on the THz range devices, as effective metamaterials circular polarizers are on demand there, but conclusions are general and can be extended to optical or microwave frequency ranges as well.

The paper is organized as following. After briefly mentioning the idea of the elliptical polarization conversion in Section II, we perform the theoretical analysis of the metamaterial polarization converter with the help of the transmission line theory in Section III. The upper limits of the conversion efficiency are derived here. We optimize and numerically characterize few examples of the circular polarizers designs in the transmission and reflection configurations in Section IV. Discussion and Conclusions Section ends up the paper.

\section{Methodology: elliptical polarization}
As we mentioned in Introduction there are two principle strategies to convert a linear polarization into an elliptical one. Assume a wave propagating along the z-axis (Fig. 1) with amplitude $E_0=1$ for simplicity. One way is to to use a bianisotropic metamaterial with two elliptical eigenpolarizations, which have different absorption or reflection coefficients (elliptical dichroism). An incident linear wave splits into elliptical eigenwaves upon the incidence on a metamaterial slab, and the eigenwave with the higher transmission coefficient will manifest the output polarization. However, such polarization converter provides at the output only a fixed specific elliptical polarization and a part of the wave intensity is lost by default in absorption or poor coupling (reflection).

\begin{figure}[htbp]
\centering
\includegraphics[width=8cm]{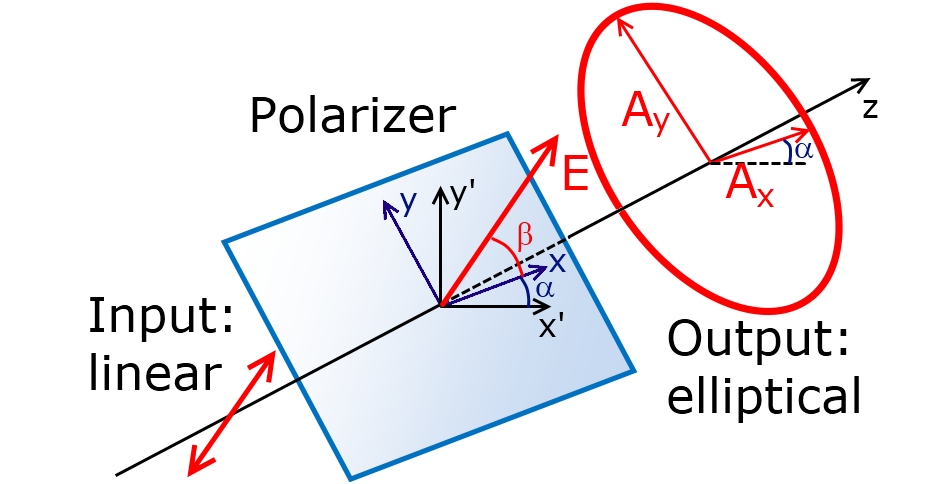}
\label{Fig1}
\caption{Operation principle of the polarization converter. An incoming linearly-polarized wave is converted into an elliptically-polarized. Rotating the linear incoming polarization with respect to polarizer (angle $\beta$) we can obtain any polarization from linear to circular. Rotating the device together with the input polarization (angle $\alpha$) we can change the orientation of the output ellipse.}
\end{figure}

Another strategy, which is more flexible, is to split the incoming wave into two nonparallel linearly polarized eigenwaves in a birefringent material. Assume the perfect transmission of both polarizations. The waves propagating with different phase velocities in the medium gain relative phase shift $\Delta\phi$. At the output port we sum them up together.  In general, an arbitrary angle between linear polarizations along with arbitrary $\Delta\phi$ (except $0$ and $180^{o}$) results in an elliptical polarization \cite{Jackson1999}. From the practical point of view, however, the most convenient case is to use two orthogonal linear polarizations $E_{x}$ and $E_{y}$ with the phase shift $\Delta\phi=90^{o}$ (see Fig. 1). Then, being combined at the output, two waves

\begin{equation}
E_{x}=A_{x}\cos(\omega t),
E_{y}=A_{y}\cos(\omega t+\Delta\phi),
\end{equation}

\noindent
give an elliptical wave described by the equation

\begin{equation}\label{Eq2}
\left(\frac{E_{x}}{A_{x}}\right)^2+\left(\frac{E_{y}}{A_{y}}\right)^2=1.
\end{equation}

The principle axes of the ellipse in Eq. (\ref{Eq2}) are along the $x-$ and $y-$directions.

Introducing angle $\beta$ between the linear incoming polarization and the $x-$axis (see Fig. 1) defines waves amplitudes $A_{x}, A_{y}$ and ellipticity $k$ (we assume perfect transmission $|t_x|=|t_y|=1$ for simplicity):

\begin{equation}
A_{x}=\cos\beta, A_{y}=\sin\beta,
k=\frac{A_x}{A_y}=\cot\beta.
\end{equation}

Changing angle $\beta$ leads to the changes in both amplitudes and ellipticity, but keeps the ellipse orientation. Rotating the polarizer and, therefore, the coordinate system $xy$ connected to it, together with the incoming wave polarization with respect to the fixed coordinate system $x'y'$ on angle $\alpha$ (see Fig. 1) allows for changing the ellipse orientation.

In such way we can obtain any elliptical polarization starting from linear ($A_x=0$ or $A_y=0$) to circular ($A_x=A_y$) with right or left rotation direction and with any ellipse orientation. We should note that in reality the transmission through a device cannot be 100\% due to imperfections, but in order to cover the whole range of possible polarizations we just need the equality of the $x-$ and $y-$polarized waves transmissions.

\section{Transmission line analysis}
The previous section gives the general strategy for obtaining any elliptical polarization out of an incoming linear one. Now we are looking for the guidelines for the polarizer design. It is desirable for the practical applications (and also fabrication-wise) to have a thin device, so we assume that our polarization converter is a thin layer of a metamaterial with thickness $H<<\lambda$.

To analyze the electromagnetic (optical) properties of the required metamaterial we employ the transmission-line (TL) approach \cite{cronin1995microwave,tretyakov2003analytical}. In the TL theory an  $x-$polarized plane wave propagating in dielectric with the refractive index $n$ is equivalent to the fundamental mode of a rectangular dielectric waveguide with the perfect electric($x-$walls) and perfect magnetic($y-$walls) boundaries. The relative (to the free-space impedance $Z_0=120\pi$ Ohm) wave impedance $\eta=\sqrt{\mu/\varepsilon}$, where $\mu$ and $\varepsilon$ are the relative magnetic permeability and electric permittivity of the material. For non-magnetic dielectrics $\mu=1$, and $\eta=1/n$. If we consider a plane interface between two non-magnetic dielectrics with refractive indices $n_{1}$ and $n_{2}$, then the reflection $r$ and transmission $t$ coefficients of the  normally incident plane wave are expressed through the wave impedances $\eta_1=1/n_{1}$ and $\eta_2=1/n_{2}$

\begin{eqnarray}
t=\frac{\frac{2}{\eta_1}}{\frac{1}{\eta_1}+\frac{1}{\eta_2}},\label{Fresnelt}\\
r=\frac{\frac{1}{\eta_1}-\frac{1}{\eta_2}}{\frac{1}{\eta_1}+\frac{1}{\eta_2}}.
\label{Fresnelr}
\end{eqnarray}

\begin{figure}[htbp]
\centering
\includegraphics[width=8cm]{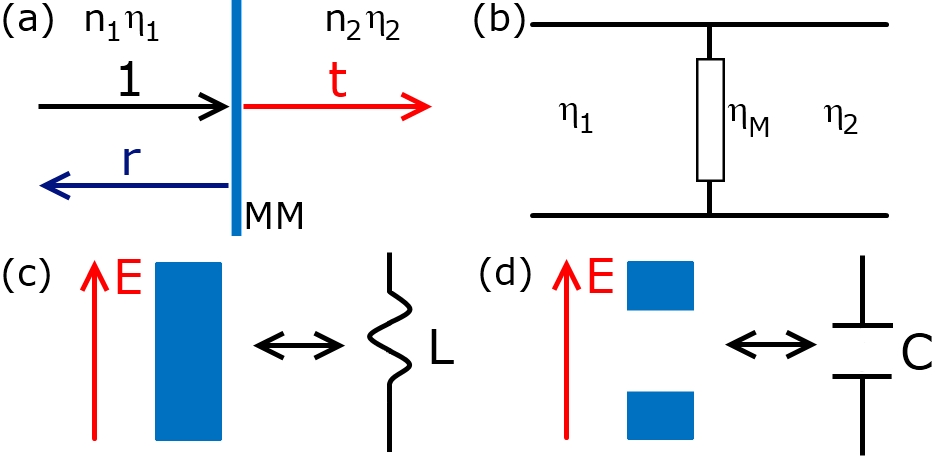}
\caption{Reflection and transmission of a plane wave incident on the metamaterial interface between two dielectrics (a) is equivalent to reflection and transmission of the wave in the connected transmission lines with different impedances $\eta_1$ and $\eta_2$ and shunted with a load $\eta_M$. In the lossless metamaterial a metallic wire along the electric field is an analogue of an inductor (c), while a dielectric gap between two metal parts is equivalent to a capacitor (d).}
\label{Fig02}
\end{figure}

If we add a thin  (much thinner than the wavelength) layer of a metamaterial with electrical impedance $Z_M$ at the interface between two dielectrics it will act as a shunt (Fig. \ref{Fig02}a). The equivalent normalized impedance of the shunt is $\eta_M=(Z_M/Z_0)(a_y/a_x)$ (Fig. \ref{Fig02}b), where $a_x, a_y$ are the periods of the unit cell of the metamaterial. The transmission $t$, reflection $r$, transmittance $T$ and reflectance $R$ of the wave normally incident from the left side are

\begin{eqnarray}
t=\frac{\frac{2}{\eta_1}}{\frac{1}{\eta_1}+\frac{1}{\eta_2}+\frac{1}{\eta_M}},\label{Fresnels+imp1}\\
r=\frac{\frac{1}{\eta_1}-\frac{1}{\eta_2}-\frac{1}{\eta_M}}{\frac{1}{\eta_1}+\frac{1}{\eta_2}+\frac{1}{\eta_M}},\\
T=\frac{\eta_1}{\eta_2}|t|^2,\\
R=|r|^2. \label{Fresnels+imp4}
\end{eqnarray}

In the extreme case of $\eta_M=\infty$ (no metamaterial layer) expressions reduce to Fresnel's formulas (\ref{Fresnelt}),(\ref{Fresnelr}). If $\eta_M=0$ (short circuit, which is equivalent to a perfectly conducting mirror), we get $t=0,~r=-1$.

To design a metamaterial-based circular polarizer we target the equal transmittances/reflectances for $x-$ and $y-$ polarizations and phase difference $\Delta\phi=90^{o}$. First, we formulate the requirements on the metamaterial impedance $\eta_M$ accordingly to formulas (\ref{Fresnels+imp1})-(\ref{Fresnels+imp4}). Then we analyze transmittance and reflectance (depending on configuration) from the point of view of maximal values which can be obtained by adjusting impedances. Then we guess the geometry to get the designated $\eta_M$. The simple rules bridging the transmission line theory and optics are that the inductive impedance requires the employment of a metallic wire along the electric field (Fig. \ref{Fig02}c), while the capacitive impedance means a dielectric gap between metallic parts (Fig. \ref{Fig02}d). Finally, the fine tuning of the design geometry is done with the help of numerical optimization in CST Microwave Studio \cite{CST}.

We begin our theoretical analysis assuming that the metamaterial is made of a perforated perfect conductor film that is a fairly good approximation for metals in the microwave and THz range. Thus, the metamaterial impedance is fully reactive (pure imaginary) $\eta_M=-ix$, where $x$ is an effective reactance. Since we use the optical notation for the time exponential $\exp(-i\omega t)$, positive $x$ means predominantly inductive impedance $\eta_M=-i\omega L$ ($x=\omega L$), while negative $x=-1/(\omega C)$ means predominantly capacitive impedance $\eta_M=i/(\omega C)$ . Then the transmission and reflection coefficients (\ref{Fresnels+imp1}-\ref{Fresnels+imp4}) are

\begin{eqnarray}
t=\frac{\frac{2}{\eta_1}}{\frac{1}{\eta_1}+\frac{1}{\eta_2}+\frac{i}{x}}=\frac{2}{1+\gamma+i\xi},\\
r=\frac{\frac{1}{\eta_1}-\frac{1}{\eta_2}-\frac{i}{x}}{\frac{1}{\eta_1}+\frac{1}{\eta_2}+\frac{i}{x}}=\frac{1-\gamma-i\xi}{1+\gamma+i\xi},\\
T=\frac{\frac{4}{\eta_1\eta_2}}{(\frac{1}{\eta_1}+\frac{1}{\eta_2})^2+\frac{1}{x^2}}=\frac{4\gamma}{(1+\gamma)^2+\xi^2},\\
R=\frac{(\frac{1}{\eta_1}-\frac{1}{\eta_2})^2+\frac{1}{x^2}}{(\frac{1}{\eta_1}+\frac{1}{\eta_2})^2+\frac{1}{x^2}}=\frac{(1-\gamma)^2+\xi^2}{(1+\gamma)^2+\xi^2}.
\end{eqnarray}

\noindent
where $\gamma=\eta_1/\eta_2=n_2/n_1$ and $\xi=\eta_1/x$.

The phases of reflected and transmitted waves are defined through the tangent functions:
\begin{eqnarray}
\label{tangentT}
\tan\phi_t=-\frac{\xi}{1+\gamma},\\
\label{tangentR}
\tan\phi_r=-\frac{2\gamma}{1-\gamma^2-\xi^2}.
\end{eqnarray}

\subsection{Metamaterial layer: transmission configuration}
Consider a single metamaterial layer in the transmission configuration \cite{Chin2008,Strikwerda2009,Roberts2012}. From the requirement of equal transmittance $T_x=T_y$ we find that the reactance for the $x-$ and $y-$polarizations should be equal in absolute values $|\xi_x|=|\xi_y|$. Among the real solutions $\xi_x=\pm \xi_y$ the $"+"$ solution leads to the same phase advance for both polarization, and, therefore, must be excluded. So, the only suitable solution is $\xi_x=-\xi_y$. That means in the simplest case that the impedance of the metamaterial should be inductive for one polarization and capacitive for another.

Consequently, the transmission phase tangents $\tan\phi_{t,x}=-\tan\phi_{t,y}$. Since the phase difference should be $90^{o}$, the product of their tangents is $\tan\phi_{t,x}\tan\phi_{t,y}=-1$. Therefore, $\tan\phi_{t,x}^2=1$ and

\begin{equation}
\tan\phi_{t,x}=-\frac{\xi_x}{1+\gamma}=\pm 1.
\end{equation}

Let us, for certainty, select the positive solution for $x-$polarization then the negative solution will correspond to $y-$ polarization, thus reactances are

\begin{equation}
\label{xi_T1}
\xi_x=-\xi_y=(1+\gamma).
\end{equation}

The transmittance is

\begin{equation}
T=\frac{2\gamma}{(1+\gamma)^2},
\end{equation}

\begin{figure}[htbp]
\centering
\includegraphics[width=8cm]{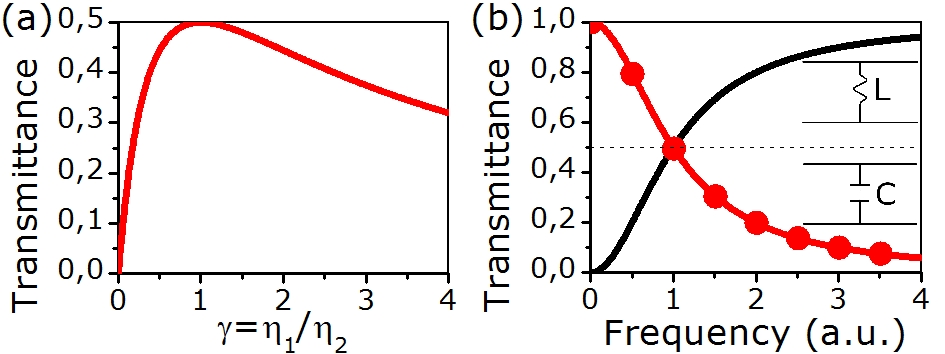}
\caption{(a) The dependence of transmittance on  $\gamma=\eta_1/\eta_2=n_2/n_1$ shows that the maximal transmittance $T=0.5$ is reached when $\gamma=1$. (b) Positive and negative $\xi$ of a metamaterial is analogous to the high and low frequency filters, respectively. Being selected in an appropriate way, they can have equal transmittances and a phase difference $90^{o}$ at the working frequency.}
\label{Fig03}
\end{figure}

\noindent
The  maximal value of transmittance $T_{max}=0.5$ can be reached for $\gamma=\eta_1/\eta_2=1$ (Fig. \ref{Fig03}a), that means the metamaterial is placed between the same surrounding materials. Physically this means either a metallic membrane suspended in air \cite{Malureanu2010} or a metamaterial coated with dielectric with the same refractive index as of the substrate.

A metamaterial satisfying this solution should exhibit inductive properties for $x-$polarization (metallic wires along the $x-$axis) and capacitive for $y-$ polarization (dielectric gapes orthogonally to the $y-$axis). In the transmission line analogy this is a high-frequency filter for the $x-$ and low-frequency filter for the $y-$polarization (Fig. \ref{Fig03}b). In the simplest case such polarization converter can consist of a set of metallic stripes or wires along the $x-$ axis.

\subsection{Metamaterial layer: reflection configuration}

Now we consider a single metamaterial layer in the reflection regime \cite{Pors2011}. Similarly to the transmission case reflectance for both linearly-polarized waves should be the same $R_x=R_y$. So, the reactances should be identical in absolute values $|\xi_x|=|\xi_y|$. Analogously to the transmission regime analysis we come to conclusion that $\xi_x=-\xi_y$, and then reactance $\xi_x$ can be extracted from the expression for the reflection phase (\ref{tangentR})

\begin{equation}
\label{EqPhirx}
\tan\phi_{r,x}=-\frac{2\xi_x}{1-\gamma^2-\xi_x^2}=\pm 1,
\end{equation}

\noindent
which is a quadratic equation for $\xi_x$. Selecting the positive sign for the $x-$polarization in equation (\ref{EqPhirx}), we obtain

\begin{equation}
\xi_x^2-2\xi_x-1+\gamma^2=0.
\end{equation}

The analysis of this quadratic equation reveals that there are no solutions for $\gamma>\sqrt{2}$, so it is not possible to reach the phase difference $90^{o}$ between polarizations with any metamaterial. For $\gamma<\sqrt{2}$ the solutions for reactances for both polarizations are

\begin{eqnarray}
\label{xiR1_1}
\xi_{x,1}=-\xi_{y,1}=1+\sqrt{2-\gamma^2},\\
\xi_{x,2}=-\xi_{y,2}=1-\sqrt{2-\gamma^2}.
\end{eqnarray}

\begin{figure}[htbp]
\centering
\includegraphics[width=4cm]{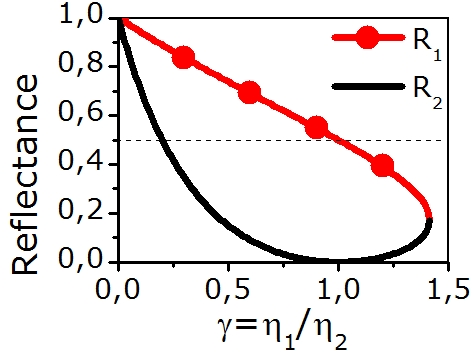}
\caption{There are two solutions for $\gamma=\eta_1/\eta_2=n_2/n_1<\sqrt{2}$. Both of them have reflectances less than 0.5 for $\gamma>1$ and can have large reflectance $R\approx 1$ for $\gamma\ll 1$. Physically this means an incidence from the material with a refractive index $n_1\gg n_2$.}
\label{Fig04}
\end{figure}

The corresponding reflectances

\begin{eqnarray}
R_1=\frac{2-\gamma+\sqrt{2-\gamma^2}}{2+\gamma+\sqrt{2-\gamma^2}},\\
R_2=\frac{2-\gamma-\sqrt{2-\gamma^2}}{2+\gamma-\sqrt{2-\gamma^2}},
\end{eqnarray}

\noindent
are shown in Fig. \ref{Fig04}. For both solutions reflectance is less than 0.5 for $1<\gamma<\sqrt{2}$. For $\gamma=1$ the first solution (\ref{xiR1_1}) coincides with the maximal transmission solution (\ref{xi_T1}). In this case the converter works simultaneously in reflection and transmission regimes, giving 50\% reflectance and 50\% transmittance.

Note, however, that the ratio $\gamma$ can be less than 1. It can happen if the second dielectric has smaller refractive index than the first one, so the incidence should occur, for example, through a high-dielectric substrate. With such impedances reflectance can be larger than 0.5, and even equal to 1 for $\gamma=0$. It is, however, not possible to use such high-dielectric substrates due to natural limitations on the material properties and also due to a poor wave in-coupling to a high-dielectric substrate from air. For a practically important case of a silicon substrate in air, $\gamma\approx 0.3$ that corresponds to the reflectance $R_1\approx 0.84$ and $R_2\approx 0.36$. So, in contrary to the transmission regime with the maximal polarization conversion up to 50\%, it is possible to achieve larger power conversions in the reflection configuration.

\subsection{Two metamaterial layers: 100\% conversion efficiency in transmission}

\begin{figure}[htbp]
\centering
\includegraphics[width=6cm]{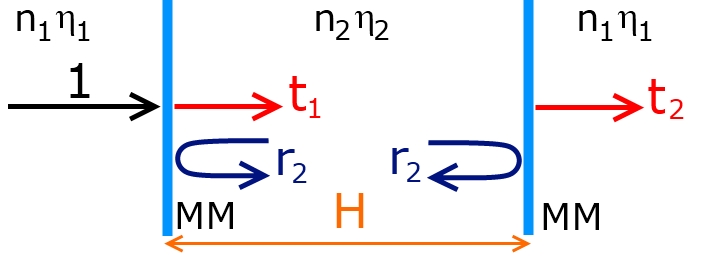}
\caption{Two identical metamaterial layers separated with a dielectric form a Fabry-Perot resonator.}
\label{Fig05}
\end{figure}

The theoretical limit for the conversion efficiency for a single layer metamaterial is 50\%. We can, however, expect larger efficiency applying two layers of metamaterial separated with a dielectric spacer \cite{Kwon2008,Strikwerda2009,Weis2009,Li2010}. Even very reflective mirrors arranged as a Fabry-Perot resonator can give transmittance close to 1. Let us consider a symmetric configuration shown in Fig. \ref{Fig05}, namely, two thin identical metamaterial layers separated with a dielectric $n_2$ and having a dielectric $n_1$ from the left and from the right.

The transmission through a Fabry-Perot resonator is \cite{Yeh1988}

\begin{equation}
t=\frac{t_1t_2\exp(i\Phi)}{1-r_2\exp(i2\Phi)},
\end{equation}

\noindent
where $\Phi=k_0n_2H$, $H$ is the thickness of the dielectric spacer between metamaterial layers, $n_2$ is its refractive index, $k_0$ is the free-space wavenumber and the meaning of the reflection and transmission coefficients $t_1, r_2, t_2$ is clear from Fig. \ref{Fig05}.

\begin{eqnarray}
t_1=\frac{2}{1+\gamma+i\xi},\\
r_2=\frac{\gamma-1-i\xi}{\gamma+1+i\xi},\\
t_2=\frac{2\gamma}{1+\gamma+i\xi}.
\end{eqnarray}

Through a tedious analysis we found that equal transmittances for the $x-$ and $y-$ polarizations occur for the same condition for the metamaterial impedance $\xi_x=-\xi_y$ as for a single layer.

The tangent of the transmission phase is

\begin{equation}
\tan\phi_t=-\frac{\xi(\gamma-\sin\Phi)}{\gamma-\frac{1}{2}(1+\gamma^2-\xi^2)\sin\Phi}.
\end{equation}
The condition for the $90^{o}$ phase difference $\tan\phi_x\tan\phi_y=-1$ leads to a quadratic equation for $\xi_x$. It has two solutions

\begin{eqnarray}
\label{xi_T2_1}
\xi_{x,1}=-\xi_{y,1}=1-\gamma\cot\Phi+\sqrt{2+\gamma^2(1+\cot^2\Phi)}~~\\
\label{xi_T2_2}
\xi_{x,2}=-\xi_{y,2}=1-\gamma\cot\Phi-\sqrt{2+\gamma^2(1+\cot^2\Phi)}~~
\end{eqnarray}

with the corresponding transmittances

\begin{eqnarray}
T_1=\frac{1}{2}\frac{1}{(\sqrt{1+2w^2}-w)^2},\\
T_2=\frac{1}{2}\frac{1}{(\sqrt{1+2w^2}+w)^2},
\end{eqnarray}

\noindent
where $w=\sin\Phi/\gamma=\frac{\eta_2}{\eta_1}\sin(k_0n_2H)=\frac{n_1}{n_2}\sin(k_0n_2H)$.

\begin{figure}[htbp]
\centering
\includegraphics[width=4cm]{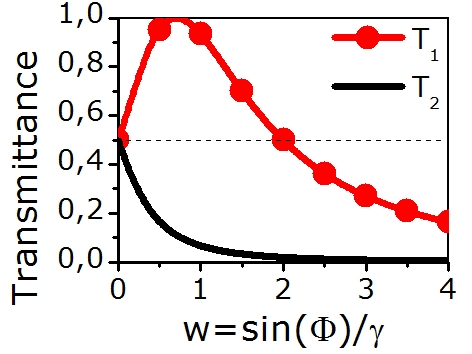}
\caption{There are two solutions for the metamaterial reactances. The first solution (\ref{xi_T2_1}) can give the transmittance up to 100\%, while the transmittance for the second solution (\ref{xi_T2_2}) cannot exceed 50\%.}
\label{Fig06}
\end{figure}

The transmittance for the second solution (Fig. \ref{Fig06}) is always $T_2<0.5$. However, choosing the first solution it is possible to achieve the full transmission: $T_1=1$ for $w=1/\sqrt{2}$. That imposes the condition for the spacer thickness $H$

\begin{equation}
\label{Eqsink0n2H}
\sin(k_0n_2H)=\gamma/\sqrt{2}=\frac{n_2}{n_1\sqrt{2}}.
\end{equation}

In the case when $n_2/n_1>\sqrt{2}$, there are no real solutions for Eq. (\ref{Eqsink0n2H}), and it is not possible to achieve even theoretically the 100\% transmittance. Such situation can happen if the metamaterials layers are separated with a high refractive index dielectric. Nevertheless, by coating the metal layers with additional dielectric (see, for example the meanderline structure in \cite{Strikwerda2009})  it is always possible to reduce the ratio $\gamma=n_2/n_1$ and improve the transmission characteristics.

\subsection{Metamaterial layer above ground plane: 100\% conversion efficiency in reflection}

\begin{figure}[htbp]
\centering
\includegraphics[width=6cm]{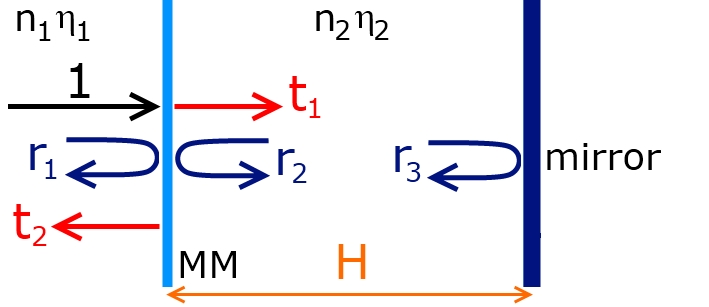}
\caption{Metamaterial layer above a ground plane has 100\% reflectance for any metamaterial impedance.}
\label{Fig07}
\end{figure}

Despite a larger conversion efficiency in the reflection configuration comparing to the transmission regime, the reflectance is not 100\% (see Subsection B). Moreover, the incidence from the high-dielectric material side requires preliminary in-coupling to a substrate and therefore some power backscattering. It is, however, possible to overcome these challenges and reach up to 100\% reflectance for the incidence from air by using a metal mirror (ground plane) below the metamaterial layer (Fig.\ref{Fig07}) \cite{Hao2007,Hao2009,Strikwerda2011,Wang2012}.

Physically such system is a kind of the Fabry-Perot resonator with a perfect mirror ($r_3=-1=\exp(i\pi)$) and a metamaterial mirror (see Fig. \ref{Fig07}). The total reflectance from this system is \cite{Yeh1988}:

\begin{equation}
\label{EqRgroundplate}
r=r_1-\frac{t_1t_2}{\exp(-i2\Phi)+r_2},
\end{equation}

\noindent
where $\Phi=k_0n_2H$ and the meaning of the reflection and transmission coefficients $r_1, t_1, r_2, t_2$ is clear from Fig. \ref{Fig07}

\begin{eqnarray}
r_1=\frac{1-\gamma-i\xi}{1+\gamma+i\xi},\\
t_1=\frac{2}{1+\gamma+i\xi},\\
r_2=\frac{\gamma-1-i\xi}{\gamma+1+i\xi},\\
t_2=\frac{2\gamma}{1+\gamma+i\xi}.
\end{eqnarray}

For the case of the perfect ground plane the reflectance is unitary $R=|r|^2=1$ for both polarizations at any frequency. The wave has simply no other options than to be reflected. A tedious analysis of the reflection phase $\phi_r$ shows that

\begin{equation}
\tan\phi_r=\frac{2(\xi+\gamma\cot\Phi)}{(\xi+\gamma\cot\Phi)^2-1}.
\end{equation}

Requirement $\tan\phi_{r,x}\tan\phi_{r,y}=-1$ ensures the $90^{o}$ phase difference between polarizations. That leads to a parametric quadratic equation

\begin{equation}
v_x^2v_y^2-v_x^2-v_y^2+4v_xv_y+1=0,
\end{equation}

\noindent
where $v_x=\xi_x+\gamma\cot\Phi, v_y=\xi_y+\gamma\cot\Phi$. This quadratic equation reduces to two independent cases of linear equations

\begin{eqnarray}
\label{Eqvxy1}
(v_x+1)(v_y-1)=-2,\\
\label{Eqvxy2}
(v_x-1)(v_y+1)=-2.
\end{eqnarray}

So, for each specific $v_x$ there are two solutions that give the required converter functionality

\begin{eqnarray}
v_{y,1}=1-\frac{2}{v_x+1},\\
v_{y,1}=-1-\frac{2}{v_x-1}.
\end{eqnarray}

It means that for any specific thickness $H$ for any given $x-$polarization reactance $\xi_x$ there exists a real-valued $y-$ polarization reactance $\xi_y$. Moreover, in contrary to all previous cases (transmission and reflection from a single metamaterial layer or transmission through a double layer), the reactances $\xi_x$ and $\xi_y$ are not obliged to be of different signs. They can be both positive or negative. It is possible to use a fully inductive (or capacitive) metamaterial for both polarizations, for example, a two-dimensional wire grid (or patches).

We note that, in principle, we have 4 variables that we can change in design: metamaterial impedances $\xi_x$ and $\xi_y$, dielectric impedances ratio $\gamma$ and the thickness of the second dielectric H. The simplest tuning can be done by the dielectric thickness $H$ optimization. An important question is whether there can be always found the value of $H$ for any given pair of $\xi_x,\xi_y$.

\begin{figure}[htbp]
\centering
\includegraphics[width=4cm]{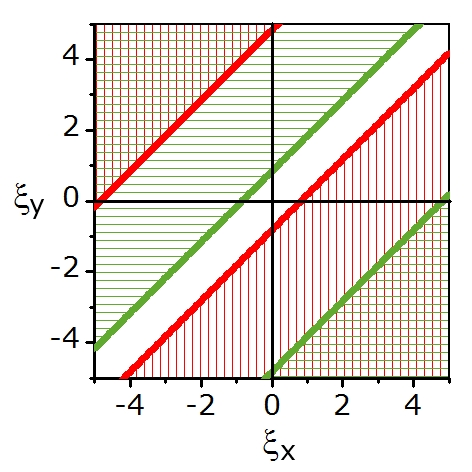}
\caption{The graphical representation of the allowed pairs of metamaterial reactances $\xi_x, \xi_y$ that give a solution to the dielectric thickness $H$. The first and second solution are marked with vertical red and horizontal green lines, respectively. There exist areas where two, one and no solutions are possible. Solutions can be of the same sign.}
\label{Fig08}
\end{figure}

For the first (\ref{Eqvxy1}) and second (\ref{Eqvxy2}) solutions we have equations for $u=\gamma\cot\Phi$

\begin{eqnarray}
(\xi_x+1+u_1)(\xi_y-1+u_1)=-2,\\
(\xi_x-1+u_2)(\xi_y+1+u_2)=-2,
\end{eqnarray}

\noindent
which have solutions

\begin{eqnarray}
u_1=\frac{1}{2}[-(\xi_x+\xi_y)\pm\sqrt{(\xi_x-\xi_y+2)^2-8}],\\
u_2=\frac{1}{2}[-(\xi_x+\xi_y)\pm\sqrt{(\xi_y-\xi_x+2)^2-8}],
\end{eqnarray}

\noindent
when conditions

\begin{eqnarray}
|\xi_x-\xi_y+2|\geq 2\sqrt{2},\\
|\xi_y-\xi_x+2|\geq 2\sqrt{2},
\end{eqnarray}

\noindent
are satisfied. The graphical representation of allowed $\xi_x, \xi_y$ is shown in Fig. \ref{Fig08}. There are values of $\xi_x, \xi_y$ corresponding to two or only one solution for $u=\gamma\cot(k_0 n_2H)$ and therefore for $H$. There are also ranges, where apparently no solutions exist. This result is expected since for the case of equal impedances $\xi_x=\xi_y$ no phase difference between polarizations can occur upon reflection.

\section{Simulation results}

In order to test the abovementioned theoretical findings we developed several polarization converters based on the same unit cell geometry, aiming the THz frequency range around 1 THz. The simulations were done in CST Microwave Studio \cite{CST}.  We applied the periodic (unit cell) boundary conditions in the $x-$ and $y-$directions, and open space boundary conditions (perfectly matched layers) in the $z-$direction. We used silver described with the Drude model ($\epsilon_\infty=5, \omega_p=1.37\times 10^{16}$rad/s, $\gamma =1.6\times 10^{13}$Hz)) as metal and silica ($n=1.5$) in most cases as dielectric. Only in the single layer reflection configuration we used silicon ($n=3.5$).

\begin{figure}[htbp]
\centering
\includegraphics[width=8cm]{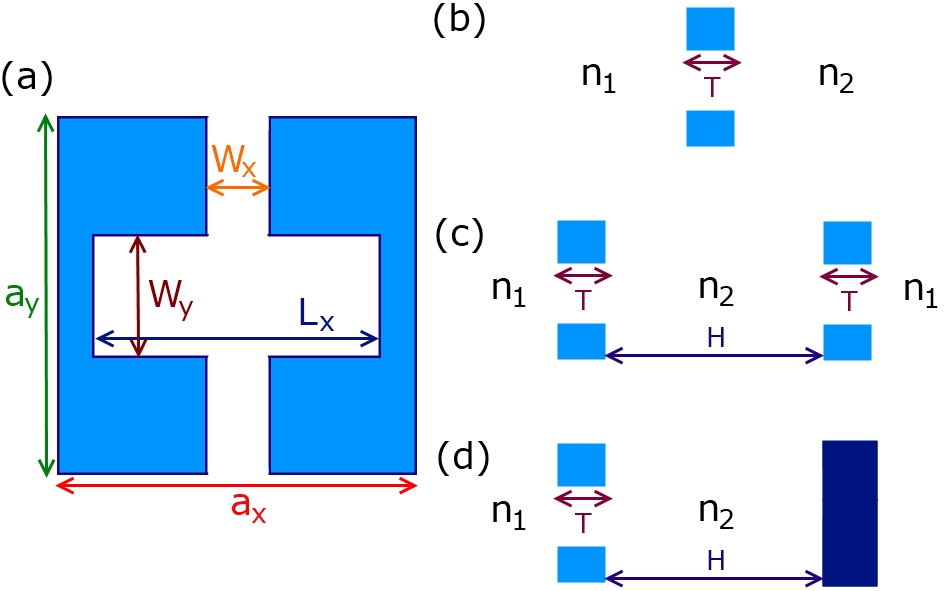}
\caption{(a) Metamaterial's unit cell design for polarization converter: view from top. Side view of the polarizer in the transmission (T1) and reflection (R1) single layer (b), transmission double layers (T2) (c) and reflection single layer with ground plane (R2) (d) configurations.}
\label{Fig09}
\end{figure}

As we mentioned before, the simplest shape for the simultaneously $x-$polarization capacitive and $y-$polarization inductive metamaterial is a stripe grid with dielectric gaps in the $x-$direction. To obtain more flexibility in the geometrical parameters we modified the stripes to the unit cell shown in Fig. \ref{Fig09}. The geometrical sizes of the characteristic features, optimized for the highest possible transmission or reflection in all four regimes are presented in Table \ref{Table}. We use the following designations: T1 - single metamaterial layer polarization converter in   transmission; R1 - single metamaterial layer polarization converter in reflection; T2 - double metamaterial layer polarization converter in transmission and R2 - single metamaterial layer polarization converter with a ground plane.

\begin{table}[ht]
\caption{Design parameters for the polarization converters}
\centering
\begin{tabular}{c c c c c}
\hline\hline
Parameter & T1 & R1 & T2 & R2 \\ [0.5ex]
\hline                  
$a_x$ ($\mu$m) & 75 & 60 & 105 & 30  \\ 
$a_y$ ($\mu$m) & 75 & 50 & 65 & 30 \\
$L_x$ ($\mu$m) & 72 & 55 & 103 & 18 \\
$W_x$ ($\mu$m) & 19 & 10 & 50 & 7.5 \\
$W_y$ ($\mu$m) & 30 & 17 & 40 & 8 \\
$T$ ($\mu$m) & 1 & 1 & 1 & 1 \\
$H$ ($\mu$m) & - & - & 30 & 55 \\ [1ex]
\hline 
\end{tabular}
\label{Table} 
\end{table}

\begin{figure}[htbp]
\centering
\includegraphics[width=8cm]{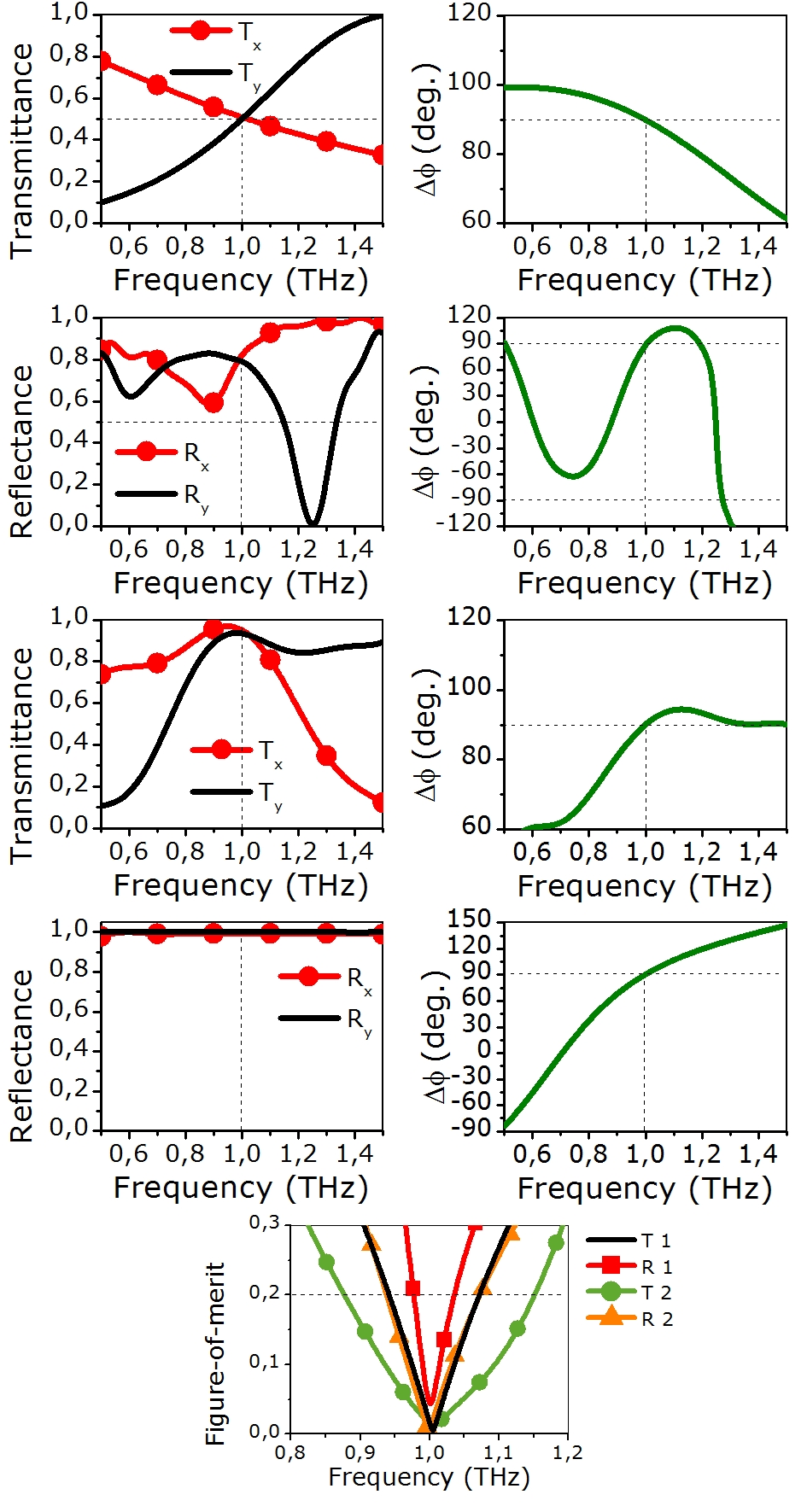}
\caption{Transmittance/reflectance and phase difference for designs (a) T1, (b) R1, (c) T2, (d) R2 are close to the theoretical maxima. (e) Figure of merit $FoM<0.2$ shows the  converter'sbandwidth, which is the largest for the transmission through two metamaterial layers (T2) design.}
\label{Fig10}
\end{figure}

The numerical results for the transmittance/reflectance and phase difference between polarizations for different designs (T1, R1, T2 and R2) are shown in Fig. \ref{Fig10}. For all designs we demonstrate the polarization conversion close to the theoretical maximum and the phase difference close to $90^{o}$. The double layer structure (T2) has the transmittance (Fig. \ref{Fig10}c) almost twice as large as for the single layer (T1) (Fig. \ref{Fig10}a). The mirror-based design R2 is, as expected, polarization insensitive and achromatic in the whole frequency range (Fig. \ref{Fig10}d).

We should mention that both the transmittance/reflectance values and the phase difference influence the conversion efficiency. To account for both influences and to determine the working bandwidth we used the approach established by Rahm et al. \cite{Weis2009}. The figure-of-merit ($FoM$) for the converter, also known as flattening, shows how close the polarization ellipse is to a circle for the input linear polarization incident at $45^{o}$. To calculate it one should get a normalized electric field vector at the output $\mathbf{E}=\hat{\textbf{x}}t_x+\hat{\textbf{y}}t_y\exp(i\Delta\phi)$, where $\hat{\textbf{x}}$ and $\hat{\textbf{y}}$ are unit vectors in the corresponding directions. Then, to calculate the angle $\psi=\arg(\mathbf{E}^2)$ and vectors corresponding to ellipse's semi-axes $\mathbf{a_1}=Re(\mathbf{E}\exp(-i\psi/2))$ and $\mathbf{a_2}=Im(\mathbf{E}\exp(-i\psi/2))$. Eventually, $FoM=1-(|\mathbf{a_2}|/|\mathbf{a_1}|)^{\pm 1}$. Plus or minus should be chosen depending on whether $\mathbf{a_1}$ or $\mathbf{a_2}$ is the major semi-axis. The figure-of-merit close to 0 shows that the polarizer works well. The working range is defined as the frequency range, where $FoM<0.2$ \cite{Weis2009}. The steepest phase difference profile is observed for the R1 design, what is reflected in the narrowest bandwidth (Fig. \ref{Fig10}(e)). The broadest FoM is observed with the T2 design due to the flatness of the phase difference spectral profile. The main parameters of the polarizers: theoretical conversion efficiency ($CE_T$), conversion efficiency extracted from simulations ($CE_S$), bandwidth ($\Delta\nu$) and relative bandwidth ($\Delta\nu_{rel}=\Delta\nu/1  $THz) are presented in Table \ref{Table2}.

\begin{table}[ht]
\caption{Parameters of the polarization converters.}
\centering
\begin{tabular}{c c c c c}
\hline\hline
Design & $CE_T$ & $CE_S$ at 1 THz & $\Delta\nu$ (GHz)& $\Delta\nu_{rel}$\\ [0.5ex]
\hline                  
T1 & 50\%& 50\% & 129 & 0.129 \\ 
R1 & 84\%& 81\% & 57 & 0.057 \\
T2 & 100\%& 94\% & 273 & 0.273 \\
R2 & 100\%& 99.7\% &134 & 0.134 \\ [1ex]
\hline 
\end{tabular}
\label{Table2} 
\end{table}

\section{Discussion and conclusions}

We have performed the general analysis of a metamaterial based polarization converter capable to transform a linearly-polarized incident wave into elliptically-polarized with any desired ellipticity and ellipse orientation. It is applicable to any thin-film (thickness $\ll$ wavelength) polarization converter with linear eigenpolarizations. We claim that a polarization converter should not necessarily be based on resonant particles (the resonance usually means an increased absorption), i.e. detuned dipoles, resonant apertures, etc., but rather on special relations between impedances for eigenpolarizations. Such impedances can correspond to the resonant or non-resonant metamaterials. The latter case usually provides a broader operation bandwidth. Even though it is not possible to remove completely parasitic capacitance and inductance, so resonances always exist. However, by a proper design we can tune the metamaterial to the frequency far away from an undesirable resonance.

As soon as the clear identification of the impedances and geometrical parameters of elements exist it is not a problem to rescale the metamaterial polarization converter to virtually any range of electromagnetic waves. In particular, we have successfully checked the scaling of the T1 design up to the telecom frequencies.

\begin{figure}[htbp]
\centering
\includegraphics[width=6cm]{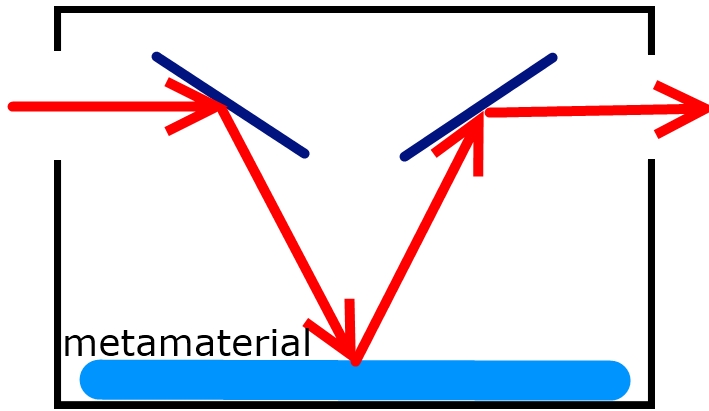}
\caption{The reflection polarizer can easily be converted into the transmission polarizer with the help of additional mirrors. For that the metamaterial layer should preserve its functionality for the incidence at small angles.}
\label{fig11}
\end{figure}

We have also showed numerically that a single type of the metamaterial unit cell can be used for any converter configuration: all four our designs T1, R1, T2, R2 are based on the same type of a unit cell differing only by optimized values of characteristic features. The one-layer transmission configuration T1 has been proved to have theoretical limit of conversion efficiency 50\%. Previously 50\% \cite{Chin2008}, 45\% \cite{Strikwerda2009} and 44\% \cite{Roberts2012} were reported. The maximal reflectance approaches 80-90\% for a single metamaterial layer with realistic dielectric properties (silicon substrate), however the steep spectral dependence of the phase shifts reduces the working range of the polarizer. Resonant antennas system was reported to provide 40\% in reflectance \cite{Pors2011}. We have demonstrated that a linear-to-circular polarization converter exhibits better performance (with the upper limit 100\% transmittance or reflectance) when two metamaterial layers are involved. The broadest working range 273 GHz is found for the transmission double layer configuration T2. The reported results for the transmittance for such systems (74\% \cite{Weis2009}, 25\% \cite{Li2010}, 80\% \cite{Strikwerda2009} and 50\% \cite{Kwon2008}) can theoretically be improved.  The case of the metamaterial above a ground plane can also be classified as double-layer system, since due to the presence of the mirror plate the wave passes twice through the metamaterial layer. Almost 100\% conversion efficiency was reported: 96\% \cite{Wang2012}, $\sim 100\%$ \cite{Strikwerda2011} and \cite{Hao2007}. The important advantage of the reflection design with a ground plate is that the requirements for the impedances values are rather softened, which can facilitate greatly the design and fabrication of the structure. Even though the transmission configuration is conventionally preferable for the real life experiments, we should note, that the reflection polarizer configuration can be easily converted into transmission one with the help of additional mirrors (Fig. \ref{fig11}).

We believe that our general approach will become a useful tool for the metamaterials based polarization converters design and development.

\section*{Acknowledgments}
The authors acknowledge P.U. Jepsen and A. Strikwerda for useful discussions. A.A. acknowledges financial support from the Danish Council for Technical and Production Sciences through the GraTer (11-116991) project.

\bibliographystyle{apsrev4-1}
\bibliography{MarkovichBibliography}

\begin{thebibliography}{35}%
\makeatletter
\providecommand \@ifxundefined [1]{%
 \@ifx{#1\undefined}
}%
\providecommand \@ifnum [1]{%
 \ifnum #1\expandafter \@firstoftwo
 \else \expandafter \@secondoftwo
 \fi
}%
\providecommand \@ifx [1]{%
 \ifx #1\expandafter \@firstoftwo
 \else \expandafter \@secondoftwo
 \fi
}%
\providecommand \natexlab [1]{#1}%
\providecommand \enquote  [1]{``#1''}%
\providecommand \bibnamefont  [1]{#1}%
\providecommand \bibfnamefont [1]{#1}%
\providecommand \citenamefont [1]{#1}%
\providecommand \href@noop [0]{\@secondoftwo}%
\providecommand \href [0]{\begingroup \@sanitize@url \@href}%
\providecommand \@href[1]{\@@startlink{#1}\@@href}%
\providecommand \@@href[1]{\endgroup#1\@@endlink}%
\providecommand \@sanitize@url [0]{\catcode `\\12\catcode `\$12\catcode
  `\&12\catcode `\#12\catcode `\^12\catcode `\_12\catcode `\%12\relax}%
\providecommand \@@startlink[1]{}%
\providecommand \@@endlink[0]{}%
\providecommand \url  [0]{\begingroup\@sanitize@url \@url }%
\providecommand \@url [1]{\endgroup\@href {#1}{\urlprefix }}%
\providecommand \urlprefix  [0]{URL }%
\providecommand \Eprint [0]{\href }%
\providecommand \doibase [0]{http://dx.doi.org/}%
\providecommand \selectlanguage [0]{\@gobble}%
\providecommand \bibinfo  [0]{\@secondoftwo}%
\providecommand \bibfield  [0]{\@secondoftwo}%
\providecommand \translation [1]{[#1]}%
\providecommand \BibitemOpen [0]{}%
\providecommand \bibitemStop [0]{}%
\providecommand \bibitemNoStop [0]{.\EOS\space}%
\providecommand \EOS [0]{\spacefactor3000\relax}%
\providecommand \BibitemShut  [1]{\csname bibitem#1\endcsname}%
\let\auto@bib@innerbib\@empty
\bibitem [{\citenamefont {Jepsen}\ \emph {et~al.}(2011)\citenamefont {Jepsen},
  \citenamefont {Cooke},\ and\ \citenamefont {Koch}}]{Jepsen2011}%
  \BibitemOpen
  \bibfield  {author} {\bibinfo {author} {\bibfnamefont {P.}~\bibnamefont
  {Jepsen}}, \bibinfo {author} {\bibfnamefont {D.}~\bibnamefont {Cooke}}, \
  and\ \bibinfo {author} {\bibfnamefont {M.}~\bibnamefont {Koch}},\ }\href
  {\doibase 10.1002/lpor.201000011} {\bibfield  {journal} {\bibinfo  {journal}
  {Laser \& Photonics Reviews}\ }\textbf {\bibinfo {volume} {5}},\ \bibinfo
  {pages} {124} (\bibinfo {year} {2011})}\BibitemShut {NoStop}%
\bibitem [{\citenamefont {Kleine-Ostmann}\ and\ \citenamefont
  {Nagatsuma}(2011)}]{Kleine-Ostmann2011}%
  \BibitemOpen
  \bibfield  {author} {\bibinfo {author} {\bibfnamefont {T.}~\bibnamefont
  {Kleine-Ostmann}}\ and\ \bibinfo {author} {\bibfnamefont {T.}~\bibnamefont
  {Nagatsuma}},\ }\href {\doibase 10.1007/s10762-010-9758-1} {\bibfield
  {journal} {\bibinfo  {journal} {Journal of Infrared, Millimeter, and
  Terahertz Waves}\ }\textbf {\bibinfo {volume} {32}},\ \bibinfo {pages} {143}
  (\bibinfo {year} {2011})}\BibitemShut {NoStop}%
\bibitem [{\citenamefont {Tonouchi}(2007)}]{Tonouchi2007}%
  \BibitemOpen
  \bibfield  {author} {\bibinfo {author} {\bibfnamefont {M.}~\bibnamefont
  {Tonouchi}},\ }\href
  {http://www.nature.com/nphoton/journal/v1/n2/abs/nphoton.2007.3.html}
  {\bibfield  {journal} {\bibinfo  {journal} {Nature Photonics}\ }\textbf
  {\bibinfo {volume} {1}},\ \bibinfo {pages} {97} (\bibinfo {year}
  {2007})}\BibitemShut {NoStop}%
\bibitem [{\citenamefont {Molter}\ \emph {et~al.}(2012)\citenamefont {Molter},
  \citenamefont {Torosyan}, \citenamefont {Ballon}, \citenamefont {Drigo},\
  and\ \citenamefont {Beigang}}]{Molter2012}%
  \BibitemOpen
  \bibfield  {author} {\bibinfo {author} {\bibfnamefont {D.}~\bibnamefont
  {Molter}}, \bibinfo {author} {\bibfnamefont {G.}~\bibnamefont {Torosyan}},
  \bibinfo {author} {\bibfnamefont {G.}~\bibnamefont {Ballon}}, \bibinfo
  {author} {\bibfnamefont {L.}~\bibnamefont {Drigo}}, \ and\ \bibinfo {author}
  {\bibfnamefont {R.}~\bibnamefont {Beigang}},\ }\href@noop {} {\bibfield
  {journal} {\bibinfo  {journal} {Optics Express}\ }\textbf {\bibinfo {volume}
  {20}},\ \bibinfo {pages} {26163} (\bibinfo {year} {2012})}\BibitemShut
  {NoStop}%
\bibitem [{\citenamefont {Masson}\ and\ \citenamefont
  {Gallot}(2006)}]{Masson2006}%
  \BibitemOpen
  \bibfield  {author} {\bibinfo {author} {\bibfnamefont {J.}~\bibnamefont
  {Masson}}\ and\ \bibinfo {author} {\bibfnamefont {G.}~\bibnamefont
  {Gallot}},\ }\href {http://www.opticsinfobase.org/abstract.cfm?id=87398}
  {\bibfield  {journal} {\bibinfo  {journal} {Optics Letters}\ }\textbf
  {\bibinfo {volume} {31}},\ \bibinfo {pages} {265} (\bibinfo {year}
  {2006})}\BibitemShut {NoStop}%
\bibitem [{\citenamefont {Arikawa}\ \emph {et~al.}(2012)\citenamefont
  {Arikawa}, \citenamefont {Wang}, \citenamefont {Belyanin},\ and\
  \citenamefont {Kono}}]{Arikawa2012}%
  \BibitemOpen
  \bibfield  {author} {\bibinfo {author} {\bibfnamefont {T.}~\bibnamefont
  {Arikawa}}, \bibinfo {author} {\bibfnamefont {X.}~\bibnamefont {Wang}},
  \bibinfo {author} {\bibfnamefont {A.~a.}\ \bibnamefont {Belyanin}}, \ and\
  \bibinfo {author} {\bibfnamefont {J.}~\bibnamefont {Kono}},\ }\href {\doibase
  10.1364/OE.20.019484} {\bibfield  {journal} {\bibinfo  {journal} {Optics
  Express}\ }\textbf {\bibinfo {volume} {20}},\ \bibinfo {pages} {19484}
  (\bibinfo {year} {2012})}\BibitemShut {NoStop}%
\bibitem [{\citenamefont {Strikwerda}\ \emph {et~al.}(2009)\citenamefont
  {Strikwerda}, \citenamefont {Fan}, \citenamefont {Tao}, \citenamefont
  {Pilon}, \citenamefont {Zhang},\ and\ \citenamefont
  {Averitt}}]{Strikwerda2009}%
  \BibitemOpen
  \bibfield  {author} {\bibinfo {author} {\bibfnamefont {A.~C.}\ \bibnamefont
  {Strikwerda}}, \bibinfo {author} {\bibfnamefont {K.}~\bibnamefont {Fan}},
  \bibinfo {author} {\bibfnamefont {H.}~\bibnamefont {Tao}}, \bibinfo {author}
  {\bibfnamefont {D.~V.}\ \bibnamefont {Pilon}}, \bibinfo {author}
  {\bibfnamefont {X.}~\bibnamefont {Zhang}}, \ and\ \bibinfo {author}
  {\bibfnamefont {R.~D.}\ \bibnamefont {Averitt}},\ }\href
  {http://www.ncbi.nlm.nih.gov/pubmed/19129881} {\bibfield  {journal} {\bibinfo
   {journal} {Optics Express}\ }\textbf {\bibinfo {volume} {17}},\ \bibinfo
  {pages} {136} (\bibinfo {year} {2009})}\BibitemShut {NoStop}%
\bibitem [{\citenamefont {Saha}\ \emph {et~al.}(2010)\citenamefont {Saha},
  \citenamefont {Ma}, \citenamefont {Grant}, \citenamefont {Khalid},\ and\
  \citenamefont {Cumming}}]{Saha2010}%
  \BibitemOpen
  \bibfield  {author} {\bibinfo {author} {\bibfnamefont {S.~C.}\ \bibnamefont
  {Saha}}, \bibinfo {author} {\bibfnamefont {Y.}~\bibnamefont {Ma}}, \bibinfo
  {author} {\bibfnamefont {J.~P.}\ \bibnamefont {Grant}}, \bibinfo {author}
  {\bibfnamefont {A.}~\bibnamefont {Khalid}}, \ and\ \bibinfo {author}
  {\bibfnamefont {D.~R.~S.}\ \bibnamefont {Cumming}},\ }\href
  {http://www.ncbi.nlm.nih.gov/pubmed/20588340} {\bibfield  {journal} {\bibinfo
   {journal} {Optics Express}\ }\textbf {\bibinfo {volume} {18}},\ \bibinfo
  {pages} {12168} (\bibinfo {year} {2010})}\BibitemShut {NoStop}%
\bibitem [{\citenamefont {Drezet}\ \emph {et~al.}(2008)\citenamefont {Drezet},
  \citenamefont {Genet},\ and\ \citenamefont {Ebbesen}}]{Drezet2008}%
  \BibitemOpen
  \bibfield  {author} {\bibinfo {author} {\bibfnamefont {A.}~\bibnamefont
  {Drezet}}, \bibinfo {author} {\bibfnamefont {C.}~\bibnamefont {Genet}}, \
  and\ \bibinfo {author} {\bibfnamefont {T.}~\bibnamefont {Ebbesen}},\ }\href
  {\doibase 10.1103/PhysRevLett.101.043902} {\bibfield  {journal} {\bibinfo
  {journal} {Physical Review Letters}\ }\textbf {\bibinfo {volume} {101}},\
  \bibinfo {pages} {1} (\bibinfo {year} {2008})}\BibitemShut {NoStop}%
\bibitem [{\citenamefont {Chin}\ \emph {et~al.}(2008)\citenamefont {Chin},
  \citenamefont {Lu},\ and\ \citenamefont {Cui}}]{Chin2008}%
  \BibitemOpen
  \bibfield  {author} {\bibinfo {author} {\bibfnamefont {J.~Y.}\ \bibnamefont
  {Chin}}, \bibinfo {author} {\bibfnamefont {M.}~\bibnamefont {Lu}}, \ and\
  \bibinfo {author} {\bibfnamefont {T.~J.}\ \bibnamefont {Cui}},\ }\href
  {\doibase 10.1063/1.3054161} {\bibfield  {journal} {\bibinfo  {journal}
  {Applied Physics Letters}\ }\textbf {\bibinfo {volume} {93}},\ \bibinfo
  {pages} {251903} (\bibinfo {year} {2008})}\BibitemShut {NoStop}%
\bibitem [{\citenamefont {Chin}\ \emph {et~al.}(2009)\citenamefont {Chin},
  \citenamefont {Gollub}, \citenamefont {Mock}, \citenamefont {Liu},
  \citenamefont {Harrison}, \citenamefont {Smith},\ and\ \citenamefont
  {Cui}}]{Chin2009}%
  \BibitemOpen
  \bibfield  {author} {\bibinfo {author} {\bibfnamefont {J.~Y.}\ \bibnamefont
  {Chin}}, \bibinfo {author} {\bibfnamefont {J.~N.}\ \bibnamefont {Gollub}},
  \bibinfo {author} {\bibfnamefont {J.~J.}\ \bibnamefont {Mock}}, \bibinfo
  {author} {\bibfnamefont {R.}~\bibnamefont {Liu}}, \bibinfo {author}
  {\bibfnamefont {C.}~\bibnamefont {Harrison}}, \bibinfo {author}
  {\bibfnamefont {D.~R.}\ \bibnamefont {Smith}}, \ and\ \bibinfo {author}
  {\bibfnamefont {T.~J.}\ \bibnamefont {Cui}},\ }\href
  {http://www.ncbi.nlm.nih.gov/pubmed/19399142} {\bibfield  {journal} {\bibinfo
   {journal} {Optics Express}\ }\textbf {\bibinfo {volume} {17}},\ \bibinfo
  {pages} {7640} (\bibinfo {year} {2009})}\BibitemShut {NoStop}%
\bibitem [{\citenamefont {Peralta}\ \emph {et~al.}(2009)\citenamefont
  {Peralta}, \citenamefont {Smirnova}, \citenamefont {Azad}, \citenamefont
  {Chen}, \citenamefont {Taylor}, \citenamefont {Brener},\ and\ \citenamefont
  {O'Hara}}]{Peralta2009}%
  \BibitemOpen
  \bibfield  {author} {\bibinfo {author} {\bibfnamefont {X.~G.}\ \bibnamefont
  {Peralta}}, \bibinfo {author} {\bibfnamefont {E.~I.}\ \bibnamefont
  {Smirnova}}, \bibinfo {author} {\bibfnamefont {A.~K.}\ \bibnamefont {Azad}},
  \bibinfo {author} {\bibfnamefont {H.-T.}\ \bibnamefont {Chen}}, \bibinfo
  {author} {\bibfnamefont {A.~J.}\ \bibnamefont {Taylor}}, \bibinfo {author}
  {\bibfnamefont {I.}~\bibnamefont {Brener}}, \ and\ \bibinfo {author}
  {\bibfnamefont {J.~F.}\ \bibnamefont {O'Hara}},\ }\href
  {http://www.ncbi.nlm.nih.gov/pubmed/19158890} {\bibfield  {journal} {\bibinfo
   {journal} {Optics Express}\ }\textbf {\bibinfo {volume} {17}},\ \bibinfo
  {pages} {773} (\bibinfo {year} {2009})}\BibitemShut {NoStop}%
\bibitem [{\citenamefont {Weis}\ \emph {et~al.}(2009)\citenamefont {Weis},
  \citenamefont {Paul}, \citenamefont {Imhof}, \citenamefont {Beigang},\ and\
  \citenamefont {Rahm}}]{Weis2009}%
  \BibitemOpen
  \bibfield  {author} {\bibinfo {author} {\bibfnamefont {P.}~\bibnamefont
  {Weis}}, \bibinfo {author} {\bibfnamefont {O.}~\bibnamefont {Paul}}, \bibinfo
  {author} {\bibfnamefont {C.}~\bibnamefont {Imhof}}, \bibinfo {author}
  {\bibfnamefont {R.}~\bibnamefont {Beigang}}, \ and\ \bibinfo {author}
  {\bibfnamefont {M.}~\bibnamefont {Rahm}},\ }\href {\doibase
  10.1063/1.3253414} {\bibfield  {journal} {\bibinfo  {journal} {Applied
  Physics Letters}\ }\textbf {\bibinfo {volume} {95}},\ \bibinfo {pages}
  {171104} (\bibinfo {year} {2009})}\BibitemShut {NoStop}%
\bibitem [{\citenamefont {Li}\ \emph {et~al.}(2010)\citenamefont {Li},
  \citenamefont {Wang}, \citenamefont {Cao}, \citenamefont {Liu},\ and\
  \citenamefont {Zhu}}]{Li2010}%
  \BibitemOpen
  \bibfield  {author} {\bibinfo {author} {\bibfnamefont {T.}~\bibnamefont
  {Li}}, \bibinfo {author} {\bibfnamefont {S.~M.}\ \bibnamefont {Wang}},
  \bibinfo {author} {\bibfnamefont {J.~X.}\ \bibnamefont {Cao}}, \bibinfo
  {author} {\bibfnamefont {H.}~\bibnamefont {Liu}}, \ and\ \bibinfo {author}
  {\bibfnamefont {S.~N.}\ \bibnamefont {Zhu}},\ }\href {\doibase
  10.1063/1.3533912} {\bibfield  {journal} {\bibinfo  {journal} {Applied
  Physics Letters}\ }\textbf {\bibinfo {volume} {97}},\ \bibinfo {pages}
  {261113} (\bibinfo {year} {2010})}\BibitemShut {NoStop}%
\bibitem [{\citenamefont {Roberts}\ and\ \citenamefont
  {Lin}(2012)}]{Roberts2012}%
  \BibitemOpen
  \bibfield  {author} {\bibinfo {author} {\bibfnamefont {A.}~\bibnamefont
  {Roberts}}\ and\ \bibinfo {author} {\bibfnamefont {L.}~\bibnamefont {Lin}},\
  }\href {http://www.ncbi.nlm.nih.gov/pubmed/22660040} {\bibfield  {journal}
  {\bibinfo  {journal} {Optics Letters}\ }\textbf {\bibinfo {volume} {37}},\
  \bibinfo {pages} {1820} (\bibinfo {year} {2012})}\BibitemShut {NoStop}%
\bibitem [{\citenamefont {Sun}\ \emph {et~al.}(2011)\citenamefont {Sun},
  \citenamefont {He}, \citenamefont {Hao},\ and\ \citenamefont
  {Zhou}}]{Sun2011}%
  \BibitemOpen
  \bibfield  {author} {\bibinfo {author} {\bibfnamefont {W.}~\bibnamefont
  {Sun}}, \bibinfo {author} {\bibfnamefont {Q.}~\bibnamefont {He}}, \bibinfo
  {author} {\bibfnamefont {J.}~\bibnamefont {Hao}}, \ and\ \bibinfo {author}
  {\bibfnamefont {L.}~\bibnamefont {Zhou}},\ }\href
  {http://www.ncbi.nlm.nih.gov/pubmed/21403731} {\bibfield  {journal} {\bibinfo
   {journal} {Optics Letters}\ }\textbf {\bibinfo {volume} {36}},\ \bibinfo
  {pages} {927} (\bibinfo {year} {2011})}\BibitemShut {NoStop}%
\bibitem [{\citenamefont {Hao}\ \emph {et~al.}(2007)\citenamefont {Hao},
  \citenamefont {Yuan}, \citenamefont {Ran}, \citenamefont {Jiang},
  \citenamefont {Kong}, \citenamefont {Chan},\ and\ \citenamefont
  {Zhou}}]{Hao2007}%
  \BibitemOpen
  \bibfield  {author} {\bibinfo {author} {\bibfnamefont {J.}~\bibnamefont
  {Hao}}, \bibinfo {author} {\bibfnamefont {Y.}~\bibnamefont {Yuan}}, \bibinfo
  {author} {\bibfnamefont {L.}~\bibnamefont {Ran}}, \bibinfo {author}
  {\bibfnamefont {T.}~\bibnamefont {Jiang}}, \bibinfo {author} {\bibfnamefont
  {J.}~\bibnamefont {Kong}}, \bibinfo {author} {\bibfnamefont {C.}~\bibnamefont
  {Chan}}, \ and\ \bibinfo {author} {\bibfnamefont {L.}~\bibnamefont {Zhou}},\
  }\href {\doibase 10.1103/PhysRevLett.99.063908} {\bibfield  {journal}
  {\bibinfo  {journal} {Physical Review Letters}\ }\textbf {\bibinfo {volume}
  {99}},\ \bibinfo {pages} {1} (\bibinfo {year} {2007})}\BibitemShut {NoStop}%
\bibitem [{\citenamefont {Hao}\ \emph {et~al.}(2009)\citenamefont {Hao},
  \citenamefont {Ren}, \citenamefont {An}, \citenamefont {Huang}, \citenamefont
  {Chen}, \citenamefont {Qiu},\ and\ \citenamefont {Zhou}}]{Hao2009}%
  \BibitemOpen
  \bibfield  {author} {\bibinfo {author} {\bibfnamefont {J.}~\bibnamefont
  {Hao}}, \bibinfo {author} {\bibfnamefont {Q.}~\bibnamefont {Ren}}, \bibinfo
  {author} {\bibfnamefont {Z.}~\bibnamefont {An}}, \bibinfo {author}
  {\bibfnamefont {X.}~\bibnamefont {Huang}}, \bibinfo {author} {\bibfnamefont
  {Z.}~\bibnamefont {Chen}}, \bibinfo {author} {\bibfnamefont {M.}~\bibnamefont
  {Qiu}}, \ and\ \bibinfo {author} {\bibfnamefont {L.}~\bibnamefont {Zhou}},\
  }\href {\doibase 10.1103/PhysRevA.80.023807} {\bibfield  {journal} {\bibinfo
  {journal} {Physical Review A}\ }\textbf {\bibinfo {volume} {80}},\ \bibinfo
  {pages} {1} (\bibinfo {year} {2009})}\BibitemShut {NoStop}%
\bibitem [{\citenamefont {Pors}\ \emph {et~al.}(2011)\citenamefont {Pors},
  \citenamefont {Nielsen}, \citenamefont {{Della Valle}}, \citenamefont
  {Willatzen}, \citenamefont {Albrektsen},\ and\ \citenamefont
  {Bozhevolnyi}}]{Pors2011}%
  \BibitemOpen
  \bibfield  {author} {\bibinfo {author} {\bibfnamefont {A.}~\bibnamefont
  {Pors}}, \bibinfo {author} {\bibfnamefont {M.~G.}\ \bibnamefont {Nielsen}},
  \bibinfo {author} {\bibfnamefont {G.}~\bibnamefont {{Della Valle}}}, \bibinfo
  {author} {\bibfnamefont {M.}~\bibnamefont {Willatzen}}, \bibinfo {author}
  {\bibfnamefont {O.}~\bibnamefont {Albrektsen}}, \ and\ \bibinfo {author}
  {\bibfnamefont {S.~I.}\ \bibnamefont {Bozhevolnyi}},\ }\href
  {http://www.ncbi.nlm.nih.gov/pubmed/21540949} {\bibfield  {journal} {\bibinfo
   {journal} {Optics Letters}\ }\textbf {\bibinfo {volume} {36}},\ \bibinfo
  {pages} {1626} (\bibinfo {year} {2011})}\BibitemShut {NoStop}%
\bibitem [{\citenamefont {Strikwerda}\ \emph {et~al.}(2011)\citenamefont
  {Strikwerda}, \citenamefont {Averitt}, \citenamefont {Fan}, \citenamefont
  {Zhang}, \citenamefont {Metcalfe},\ and\ \citenamefont
  {Wraback}}]{Strikwerda2011}%
  \BibitemOpen
  \bibfield  {author} {\bibinfo {author} {\bibfnamefont {A.~C.}\ \bibnamefont
  {Strikwerda}}, \bibinfo {author} {\bibfnamefont {R.~D.}\ \bibnamefont
  {Averitt}}, \bibinfo {author} {\bibfnamefont {K.}~\bibnamefont {Fan}},
  \bibinfo {author} {\bibfnamefont {X.}~\bibnamefont {Zhang}}, \bibinfo
  {author} {\bibfnamefont {G.~D.}\ \bibnamefont {Metcalfe}}, \ and\ \bibinfo
  {author} {\bibfnamefont {M.}~\bibnamefont {Wraback}},\ }\href {\doibase
  10.1142/S0129156411006878} {\bibfield  {journal} {\bibinfo  {journal}
  {International Journal of High Speed Electronics and Systems (IJHSES)}\
  }\textbf {\bibinfo {volume} {20}},\ \bibinfo {pages} {583} (\bibinfo {year}
  {2011})}\BibitemShut {NoStop}%
\bibitem [{\citenamefont {Wang}\ \emph {et~al.}(2012)\citenamefont {Wang},
  \citenamefont {Chakrabarty}, \citenamefont {Minkowski}, \citenamefont {Sun},\
  and\ \citenamefont {Wei}}]{Wang2012}%
  \BibitemOpen
  \bibfield  {author} {\bibinfo {author} {\bibfnamefont {F.}~\bibnamefont
  {Wang}}, \bibinfo {author} {\bibfnamefont {A.}~\bibnamefont {Chakrabarty}},
  \bibinfo {author} {\bibfnamefont {F.}~\bibnamefont {Minkowski}}, \bibinfo
  {author} {\bibfnamefont {K.}~\bibnamefont {Sun}}, \ and\ \bibinfo {author}
  {\bibfnamefont {Q.-H.}\ \bibnamefont {Wei}},\ }\href {\doibase
  10.1063/1.4731792} {\bibfield  {journal} {\bibinfo  {journal} {Applied
  Physics Letters}\ }\textbf {\bibinfo {volume} {101}},\ \bibinfo {pages}
  {023101} (\bibinfo {year} {2012})}\BibitemShut {NoStop}%
\bibitem [{\citenamefont {Singh}\ \emph {et~al.}(2010)\citenamefont {Singh},
  \citenamefont {Plum}, \citenamefont {Zhang},\ and\ \citenamefont
  {Zheludev}}]{Singh2010}%
  \BibitemOpen
  \bibfield  {author} {\bibinfo {author} {\bibfnamefont {R.}~\bibnamefont
  {Singh}}, \bibinfo {author} {\bibfnamefont {E.}~\bibnamefont {Plum}},
  \bibinfo {author} {\bibfnamefont {W.}~\bibnamefont {Zhang}}, \ and\ \bibinfo
  {author} {\bibfnamefont {N.~I.}\ \bibnamefont {Zheludev}},\ }\href
  {http://www.ncbi.nlm.nih.gov/pubmed/20588473} {\bibfield  {journal} {\bibinfo
   {journal} {Optics Express}\ }\textbf {\bibinfo {volume} {18}},\ \bibinfo
  {pages} {13425} (\bibinfo {year} {2010})}\BibitemShut {NoStop}%
\bibitem [{\citenamefont {Li}\ \emph {et~al.}(2011)\citenamefont {Li},
  \citenamefont {Yang}, \citenamefont {Wang},\ and\ \citenamefont
  {Zhao}}]{Li2011}%
  \BibitemOpen
  \bibfield  {author} {\bibinfo {author} {\bibfnamefont {S.}~\bibnamefont
  {Li}}, \bibinfo {author} {\bibfnamefont {Z.}~\bibnamefont {Yang}}, \bibinfo
  {author} {\bibfnamefont {J.}~\bibnamefont {Wang}}, \ and\ \bibinfo {author}
  {\bibfnamefont {M.}~\bibnamefont {Zhao}},\ }\href
  {http://www.ncbi.nlm.nih.gov/pubmed/21200407} {\bibfield  {journal} {\bibinfo
   {journal} {Journal of the Optical Society of America. A}\ }\textbf {\bibinfo
  {volume} {28}},\ \bibinfo {pages} {19} (\bibinfo {year} {2011})}\BibitemShut
  {NoStop}%
\bibitem [{\citenamefont {Mutlu}\ \emph {et~al.}(2011)\citenamefont {Mutlu},
  \citenamefont {Akosman}, \citenamefont {Serebryannikov},\ and\ \citenamefont
  {Ozbay}}]{Mutlu2011}%
  \BibitemOpen
  \bibfield  {author} {\bibinfo {author} {\bibfnamefont {M.}~\bibnamefont
  {Mutlu}}, \bibinfo {author} {\bibfnamefont {A.~E.}\ \bibnamefont {Akosman}},
  \bibinfo {author} {\bibfnamefont {A.~E.}\ \bibnamefont {Serebryannikov}}, \
  and\ \bibinfo {author} {\bibfnamefont {E.}~\bibnamefont {Ozbay}},\ }\href
  {http://www.ncbi.nlm.nih.gov/pubmed/21540958} {\bibfield  {journal} {\bibinfo
   {journal} {Optics Letters}\ }\textbf {\bibinfo {volume} {36}},\ \bibinfo
  {pages} {1653} (\bibinfo {year} {2011})}\BibitemShut {NoStop}%
\bibitem [{\citenamefont {Zhao}\ \emph {et~al.}(2012)\citenamefont {Zhao},
  \citenamefont {Belkin},\ and\ \citenamefont {Al\`{u}}}]{Zhao2012}%
  \BibitemOpen
  \bibfield  {author} {\bibinfo {author} {\bibfnamefont {Y.}~\bibnamefont
  {Zhao}}, \bibinfo {author} {\bibfnamefont {M.~A.}\ \bibnamefont {Belkin}}, \
  and\ \bibinfo {author} {\bibfnamefont {A.}~\bibnamefont {Al\`{u}}},\ }\href
  {http://dx.doi.org/10.1038/ncomms1877} {\bibfield  {journal} {\bibinfo
  {journal} {Nature Communications}\ }\textbf {\bibinfo {volume} {3}},\
  \bibinfo {pages} {870} (\bibinfo {year} {2012})}\BibitemShut {NoStop}%
\bibitem [{\citenamefont {Sabah}\ and\ \citenamefont
  {Roskos}(2012)}]{Sabah2012}%
  \BibitemOpen
  \bibfield  {author} {\bibinfo {author} {\bibfnamefont {C.}~\bibnamefont
  {Sabah}}\ and\ \bibinfo {author} {\bibfnamefont {H.}~\bibnamefont {Roskos}},\
  }\href@noop {} {\bibfield  {journal} {\bibinfo  {journal} {Progres in
  Electromagnetics Research}\ }\textbf {\bibinfo {volume} {124}},\ \bibinfo
  {pages} {301} (\bibinfo {year} {2012})}\BibitemShut {NoStop}%
\bibitem [{\citenamefont {Gansel}\ \emph {et~al.}(2012)\citenamefont {Gansel},
  \citenamefont {Latzel}, \citenamefont {Frölich}, \citenamefont {Kaschke},
  \citenamefont {Thiel},\ and\ \citenamefont {Wegener}}]{Gansel2012}%
  \BibitemOpen
  \bibfield  {author} {\bibinfo {author} {\bibfnamefont {J.~K.}\ \bibnamefont
  {Gansel}}, \bibinfo {author} {\bibfnamefont {M.}~\bibnamefont {Latzel}},
  \bibinfo {author} {\bibfnamefont {A.}~\bibnamefont {Frölich}}, \bibinfo
  {author} {\bibfnamefont {J.}~\bibnamefont {Kaschke}}, \bibinfo {author}
  {\bibfnamefont {M.}~\bibnamefont {Thiel}}, \ and\ \bibinfo {author}
  {\bibfnamefont {M.}~\bibnamefont {Wegener}},\ }\href {\doibase
  10.1063/1.3693181} {\bibfield  {journal} {\bibinfo  {journal} {Applied
  Physics Letters}\ }\textbf {\bibinfo {volume} {100}},\ \bibinfo {pages}
  {101109} (\bibinfo {year} {2012})}\BibitemShut {NoStop}%
\bibitem [{\citenamefont {Caloz}\ and\ \citenamefont {Itoh}(2006)}]{Caloz2006}%
  \BibitemOpen
  \bibfield  {author} {\bibinfo {author} {\bibfnamefont {C.}~\bibnamefont
  {Caloz}}\ and\ \bibinfo {author} {\bibfnamefont {T.}~\bibnamefont {Itoh}},\
  }\href@noop {} {\emph {\bibinfo {title} {Electromagnetic metamaterials:
  transmission line theory and microwave applications: the engineering
  approach}}}\ (\bibinfo  {publisher} {Wiley-IEEE Press},\ \bibinfo {year}
  {2006})\BibitemShut {NoStop}%
\bibitem [{\citenamefont {Jackson}(1999)}]{Jackson1999}%
  \BibitemOpen
  \bibfield  {author} {\bibinfo {author} {\bibfnamefont {J.}~\bibnamefont
  {Jackson}},\ }\href@noop {} {\emph {\bibinfo {title} {Classical
  electrodynamics}}},\ Vol.~\bibinfo {volume} {67}\ (\bibinfo  {publisher}
  {John Wiley \& Sons, New York},\ \bibinfo {year} {1999})\ p.\ \bibinfo
  {pages} {841}\BibitemShut {NoStop}%
\bibitem [{\citenamefont {Cronin}(1995)}]{cronin1995microwave}%
  \BibitemOpen
  \bibfield  {author} {\bibinfo {author} {\bibfnamefont {N.~J.}\ \bibnamefont
  {Cronin}},\ }\href@noop {} {\emph {\bibinfo {title} {Microwave and optical
  waveguides}}}\ (\bibinfo  {publisher} {Taylor \& Francis},\ \bibinfo {year}
  {1995})\BibitemShut {NoStop}%
\bibitem [{\citenamefont {Tretyakov}(2003)}]{tretyakov2003analytical}%
  \BibitemOpen
  \bibfield  {author} {\bibinfo {author} {\bibfnamefont {S.}~\bibnamefont
  {Tretyakov}},\ }\href@noop {} {\emph {\bibinfo {title} {Analytical modeling
  in applied electromagnetics}}}\ (\bibinfo  {publisher} {Artech House
  Publishers},\ \bibinfo {year} {2003})\BibitemShut {NoStop}%
\bibitem [{\citenamefont {CST}()}]{CST}%
  \BibitemOpen
  \bibfield  {author} {\bibinfo {author} {\bibnamefont {CST}},\ }\href
  {http://cst.com} {\enquote {\bibinfo {title} {Computer simulation technology,
  as},}\ }\BibitemShut {NoStop}%
\bibitem [{\citenamefont {Malureanu}\ \emph {et~al.}(2010)\citenamefont
  {Malureanu}, \citenamefont {Jepsen}, \citenamefont {S.}, \citenamefont {L.},
  \citenamefont {G.}, \citenamefont {A.},\ and\ \citenamefont
  {V.}}]{Malureanu2010}%
  \BibitemOpen
  \bibfield  {author} {\bibinfo {author} {\bibfnamefont {R.}~\bibnamefont
  {Malureanu}}, \bibinfo {author} {\bibfnamefont {P.~U.}\ \bibnamefont
  {Jepsen}}, \bibinfo {author} {\bibfnamefont {X.}~\bibnamefont {S.}}, \bibinfo
  {author} {\bibfnamefont {Z.}~\bibnamefont {L.}}, \bibinfo {author}
  {\bibfnamefont {C.~D.}\ \bibnamefont {G.}}, \bibinfo {author} {\bibfnamefont
  {A.}~\bibnamefont {A.}}, \ and\ \bibinfo {author} {\bibfnamefont {L.~A.}\
  \bibnamefont {V.}},\ }\href@noop {} {\bibfield  {journal} {\bibinfo
  {journal} {Proceedings SPIE}\ }\textbf {\bibinfo {volume} {7711}},\ \bibinfo
  {pages} {77110M} (\bibinfo {year} {2010})}\BibitemShut {NoStop}%
\bibitem [{\citenamefont {Kwon}\ \emph {et~al.}(2008)\citenamefont {Kwon},
  \citenamefont {Werner},\ and\ \citenamefont {Werner}}]{Kwon2008}%
  \BibitemOpen
  \bibfield  {author} {\bibinfo {author} {\bibfnamefont {D.-H.}\ \bibnamefont
  {Kwon}}, \bibinfo {author} {\bibfnamefont {P.~L.}\ \bibnamefont {Werner}}, \
  and\ \bibinfo {author} {\bibfnamefont {D.~H.}\ \bibnamefont {Werner}},\
  }\href {http://www.ncbi.nlm.nih.gov/pubmed/18679452} {\bibfield  {journal}
  {\bibinfo  {journal} {Optics Express}\ }\textbf {\bibinfo {volume} {16}},\
  \bibinfo {pages} {11802} (\bibinfo {year} {2008})}\BibitemShut {NoStop}%
\bibitem [{\citenamefont {Yeh}(1988)}]{Yeh1988}%
  \BibitemOpen
  \bibfield  {author} {\bibinfo {author} {\bibfnamefont {P.}~\bibnamefont
  {Yeh}},\ }\href@noop {} {\emph {\bibinfo {title} {Optical waves in layered
  media}}},\ Vol.~\bibinfo {volume} {95}\ (\bibinfo  {publisher} {Wiley Online
  Library},\ \bibinfo {year} {1988})\BibitemShut {NoStop}%
\end{thebibliography}%


%

\end{document}